\documentclass[12pt,a4paper]{iopart}

\expandafter\let\csname equation*\endcsname=\relax
\expandafter\let\csname endequation*\endcsname=\relax

\usepackage{hyperref}
\usepackage{graphicx}
\usepackage{amsmath,amsfonts,amsthm}

\newcommand{\beq}{\begin{equation}}
\newcommand{\eeq}{\end{equation}}

\numberwithin{equation}{section}

\begin{document}
\title{Numerical simulation of a lattice polymer model at its integrable point}
\author{A Bedini$^1$, A L Owczarek$^1$ and T Prellberg$^2$}
\address{$^1$ Department of Mathematics and Statistics,
  The University of Melbourne, Parkville, Vic 3010, Australia.}
\address{$^2$ School of Mathematical Sciences, Queen Mary University
  of London Mile End Road, London E1 4NS, UK.}
\ead{abedini@ms.unimelb.edu.au, owczarek@unimelb.edu.au, t.prellberg@qmul.ac.uk}

\begin{abstract}
  We revisit an integrable lattice model of polymer collapse using numerical simulations. This model was first studied by Bl\"ote and Nienhuis in J. Phys. A. {\bf 22}, 1415 (1989) and it describes polymers with some attraction, providing thus a model for the polymer collapse transition. At a particular set of Boltzmann weights the model is integrable and the exponents $\nu=12/23\approx 0.522$ and $\gamma=53/46\approx 1.152$ have been computed via identification of the scaling dimensions $x_t=1/12$ and $x_h=-5/48$. We directly  investigate the polymer scaling exponents via Monte Carlo simulations using the PERM algorithm. By simulating this polymer model for walks up to length $4096$ we find $\nu=0.576(6)$ and $\gamma=1.045(5)$, which are clearly different from the predicted values. Our estimate for the exponent $\nu$ is compatible with the known $\theta$-point value of $4/7$ and in agreement with very recent numerical evaluation by Foster and Pinettes \cite{foster2012a-a}.
\end{abstract}

\maketitle

\section{Introduction}
\label{sec:introduction}

The study of the critical properties of lattice polymers, and thus of $O(n)$ models when we let $n \to 0$, in two dimensions has been ongoing over decades theoretically and numerically. Nienhuis, in 1982 \cite{nienhuis1982a-a}, considered a model of non-intersecting loops on the hexagonal lattice which allowed him to compute the critical exponents for free self-avoiding walks ($\nu = 3/4$, $\gamma = 43/23$), which model dilute polymers, and for dense polymers ($\nu = 1/2$, $\gamma = 19/16$). In 1987 Duplantier and Saleur \cite{duplantier1987a-a} were able to model bond interactions introducing vacancies on the same lattice, obtaining a full set of critical exponents for the polymer collapse transition in the interacting self-avoiding walk (ISAW) model. This collapse transition, a tri-critical point, goes under the name of `$\theta$-point' and has critical exponents $\nu = 4/7$ and $\gamma = 8/7$. There was debate at the time over the surface exponents which was resolved by Vanderzande et al. \cite{Vanderzande:1991bv} and Stella et. al \cite{Stella:1993fz} on the hexagonal lattice and on the square lattice by Foster et al. \cite{Foster:1992gc}.

In the quest for a solvable $O(n)$ model on the square lattice, Bl\"ote and Nienhuis in 1989 \cite{blote1989a-a} considered a lattice model (related to the Izergin-Korepin vertex model) which includes weights for site-collisions and straight segments (stiffness). For this model five critical branches are exactly known \cite{blote1989a-a, batchelor1989a-a,nienhuis1990a-a,warnaar1992b-a}. In one of these branches (named `branch 0' in \cite{blote1989a-a}) straight segments are completely suppressed and it can be shown that in this case the model maps to the ISAW model on the Manhattan lattice for which the conjectured exponents are $\nu = 4/7$, $\gamma = 6/7$ \cite{Bradley:1990eh, batchelor1993a-a, prellberg1994a-:a}. Two other branches correspond to dense and dilute polymers as obtained by Nienhuis in \cite{nienhuis1982a-a}, and the two remaining branches are, respectively, associated with a combination of Ising-like and $O(n)$ critical behaviour and with a new tri-critical point. This other tri-critical point, which we shall refer as a the BN-point, is another candidate for describing a collapsing polymer and has exponents $\nu = 12/23$ and $\gamma = 53/46$  \cite{warnaar1992b-a}. The configurations associated with this particular $O(n)$ model, which we shall call Vertex-Interacting Self-Avoiding Walks (VISAWs), are forbidden to cross and therefore are a subset of self-avoiding trails. The Boltzmann weights corresponding to BN-point are known exactly and can be expressed as algebraic numbers. 

Foster and Pinettes \cite{foster2003a-a} have studied the semi-flexible VISAW at this special BN-point, and have also studied the VISAW model without stiffness, using the corner transfer matrix renomalisation group method. Some agreement and some discrepancy with the scaling dimensions proposed \cite{blote1989a-a, batchelor1989a-a,nienhuis1990a-a,warnaar1992b-a} was found in \cite{foster2003a-a} and a first order nature to the transition was conjectured. Very recently Foster and Pinettes \cite{foster2012a-a} have used transfer matrices and the Density Matrix Renormalisation Group Method (DMRG) to consider the bulk and surface exponents of these models. They have found values of the exponent $\nu$ much closer to $4/7$ than $12/23$. In this paper we study by means of Monte Carlo simulation the semi-flexible VISAW polymer model precisely at the BN-point. We find estimates for the exponents, and hence the scaling dimensions, that are in harmony with those found by Foster and Pinettes \cite{foster2012a-a}  and at variance with those predicted by Warnaar {\it et al.\ }\cite{warnaar1992b-a} .

\section{Semi-flexible VISAW}
\label{sec:semi-flex-isat}

\begin{figure}[ht!]
  \centering
  \includegraphics{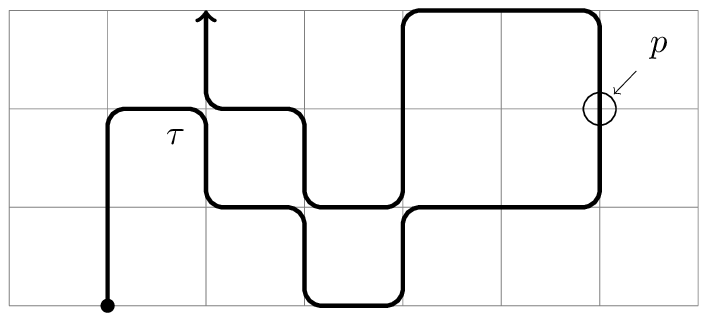}
  \caption{An example of a semi-flexible VISAW configuration with three
    ($m = 3$) collisions associated with a
    Boltzmann weight $\tau$ and five ($s = 5$) straight segments, each
    associated with a Boltzmann weight $p$.}
  \label{fig:sfVISAW}
\end{figure}

The semi-flexible VISAW model can be defined as follows. Self-avoiding
trails, or simply trails, are lattice paths that can be formed such
that they never visit the same bond more than once. Such paths can
generally visit the same site of the lattice either by a
\emph{collision}, where the trail touches itself, or via a \emph{crossing}, where two
straight segments of the path cross over one another.  Consider the
\emph{subset} of bond-avoiding lattice paths (trails) on the square
lattice, $\mathcal V_n$, where \emph{no} crossings are allowed.  Given
such a restricted trail $\psi_n \in \mathcal V_n$, we associate an
energy $-\varepsilon_t$ every time the path visits the same site more
than once, which it can only do by \emph{colliding} with itself:
see Figure~\ref{fig:sfVISAW}. Additionally, we define a straight
segment of the trail by two consecutive parallel edges, and we
associate an energy $-\varepsilon_s$ to each straight segment of the
trail, modelling the stiffness of the polymer chain.

For each configuration $\psi_n \in \mathcal V_n$ we count the number
$m(\psi_n)$ of doubly-visited sites and $s(\psi_n)$ of straight
segments: see Figure~\ref{fig:sfVISAW}. Hence we associate with
each configuration a Boltzmann weight $\tau^{m(\psi_n)} p^{s(\psi_n)}$
where $\tau = \exp(\beta \varepsilon_t)$, $p = \exp(\beta
\varepsilon_s)$, and $\beta$ is the inverse temperature $1/k_B T$. The
partition function of the model is given by
\begin{equation}
  \label{eq:canonical}
  Z_n(\tau, p) = \sum_{\psi_n\in\mathcal V_n}\
  \tau^{m(\psi_n)} p^{s(\psi_n)}
  .
\end{equation}
The finite-length reduced free energy is
\begin{equation}
	\kappa_n(T) = \frac{1}{n} \log\ Z_n
\end{equation}
and the thermodynamic limit is obtained by taking the limit of large $n$, i.e.,
\begin{equation}
	\kappa(T) = \lim_{n \to \infty} \kappa_n(T).
\end{equation}
It is expected that there is a collapse phase transition at a
temperature $T_c$ characterized by a non-analyticity in $\kappa(T)$. Equivalently, one can think of  varying $\tau$ at fixed $p$ so that there is a collapse at some value of $\tau=\tau_c(p)$.

The probability of a configuration $\psi_n$ is then
\begin{equation}
  p(\psi_n; \tau, p) = \frac{ \tau^{m(\psi_n)}
    p^{s(\psi_n)} }{ Z_n(\tau, p) }
  ,
\end{equation}
and the average of any quantity $Q$ over the ensemble set of path
$\mathcal V_n$ is given generically by
\begin{equation}
  \langle Q \rangle_n(\tau, p) = \sum_{\psi_n\in\mathcal
    V_n} Q(\psi_n) \, p(\psi_n; \tau, p)
  .
\end{equation}

In this paper we are interested in the following quantities. We calculate three measures of the size of the polymer, $\langle R_e^2 \rangle_n $, $\langle R_m^2 \rangle_n $ and $\langle R_g^2 \rangle_n $, defined as follows. We specify any $n$-step path $\psi_n$ on a lattice by a
sequence ${{\bf r}_0, {\bf r}_1, \ldots, {\bf r}_n}$ of vector positions
of the vertices of that path.
Firstly, we are interested in the average-square end-to-end distance
\begin{equation}
\langle R_e^2 \rangle_n = \langle {\bf r}_n \cdot {\bf r}_n \rangle\; ,
\end{equation}
secondly, the ensemble average of the mean-square distance of a monomer from the
endpoints
\begin{equation}
\langle R_m^2 \rangle_n = \frac{1}{n+1} \sum_{i=0}^{n} \langle {\bf
r}_i \cdot {\bf r}_i \rangle \; ,
\end{equation}
and defining the average centre-of-mass as
\begin{equation}
\langle R_c^2 \rangle_n = \frac{1}{(n+1)^2} \sum_{i=0}^{n}
\sum_{j=0}^{n} \langle {\bf r}_i \cdot {\bf r}_j \rangle \; ,
\end{equation}
we, thirdly, are interested in the average radius-of-gyration
\begin{equation}
\langle R_g^2 \rangle_n = \langle R_m^2 \rangle_n - \langle R_c^2
\rangle_n\;.
\end{equation}
In the above formulae we use ${\bf r}_0 \equiv {\bf0}$.

\subsection{Scaling}
\label{subsec:scaling}

The partition function at, and above, the collapse temperature is
believed to scale as
\begin{equation}
  \label{eq:scaling-z}
  Z_n \sim D \mu^n n^{\gamma-1}
\end{equation}
where $\mu$ is known as the connective `constant' and is related to
the thermodynamic free energy via
\begin{equation}
\label{eq:mu-def}
  \mu = e^{\kappa(T)}\;.
\end{equation}
The constant $D$ is also temperature dependent but $\gamma$ is
expected to be universal, depending only on the temperature in as much
as its value is above, or at, the collapse temperature. In two
dimensions it is well established \cite{nienhuis1982a-a} that for
$T>T_c$ (equivalently $\tau < \tau_c(p)$) we have $\gamma = 43/32$.

The collapse transition can also be characterized via a change in the
scaling of the size of the polymer with temperature. The three
measures of the size of the polymer defined above are expected to
scale as
\begin{equation}
  R^2_n \sim C_R n^{2\nu}\; ,
\end{equation}
where the amplitude $C_R$ is non-universal and temperature dependent,
while $\nu$ is expected to be universal, depending only on the
temperature in as much as its value is above, at, or below, the
collapse point.  In two dimensions it is also estabslished
\cite{nienhuis1982a-a} that $\nu=3/4$ for $T>T_c$.

Duplantier and Saleur \cite{duplantier1987a-a} identified a
tri-critical point, known as the $\theta$-point, which is  expected to
describe the collapse of a polymer in two dimensions. This point has
thermal and magnetic scaling dimensions $x_t = 1/4$, $x_h = 0$ and consequently polymer
exponents $\nu = 4/7$ and $\gamma = 8/7$. On the other hand, Warnaar {\it et al.\ }\cite{warnaar1992b-a}  predicted 
$\nu=12/23$ and $\gamma=53/46$  for the semi-flexible VISAW model at its collapse point.

\subsection{Amplitudes}
\label{sec:amplitudes}

One can also usefully define the finite-length amplitude ratios
\begin{equation}
  A_n = \frac{\langle R_g^2 \rangle_n}{\langle R_e^2 \rangle_n}
  \quad \text{ and } \quad
  B_n = \frac{\langle R_m^2 \rangle_n}{\langle R_e^2 \rangle_n}
  \; ,
\end{equation}
since these approach universal values \cite{madras1988a-a}
\begin{equation}
  A_n \rightarrow A_\infty=\frac{C_{R_g}}{C_{R_e}}
  \quad \text{ and } \quad
  B_n \rightarrow B_\infty=\frac{C_{R_m}}{C_{R_e}}
\end{equation}
in the limit $n\rightarrow \infty$. For collapsing polymers, the limiting
values should depend only on dimension and whether the temperature is
above or at the collapse transition point.

For free self-avoiding walks (which should include the VISAW model at
high temperatures) it was predicted
\cite{cardy1989a-a,caracciolo1990a-a} that
\begin{equation}
  \lambda A_\infty - 2 B_\infty + \frac{1}{2}=0\;.
\label{eqn:identity}
\end{equation}
In the derivation \cite{cardy1989a-a} of this invariant the factor
multiplying $A_\infty$ was given by
\begin{equation}
  \lambda = 2 + \frac{y_t}{y_h}\; ,
\end{equation}
where $y_t=4/3$ and $y_h=91/48$ are the thermal and magnetic
renormalisation group eigenvalues, respectively, of the dilute O(0)
model. These eigenvalues are related to the conformal scaling
dimensions via $y=2-x$. Hence
\begin{equation}
  \lambda (x_t,x_h) = 2 + \frac{2-x_t}{2-x_h}\; .
\end{equation}
The identity (\ref{eqn:identity}) implies that one can estimate this function of the
scaling dimensions as
\begin{equation}
\lambda (x_t,x_h) = \frac{4 B_\infty-1}{2A_\infty}\;,
\end{equation}
from estimates of $A_\infty$ and $B_\infty$.
This was done for various collapse models in \cite{owczarek1994b-:a}. 
Hence, if we have a conjectured value of one of the scaling
dimensions we can estimate the other from an estimate of
$\lambda$. 

\section{Integrable Bl\"{o}te-Nienhuis Point}
\label{sec:bn-point}

The special multi-critical point of the $O(n)$ model that maps to the semi-
flexible VISAW and allows for the calculation  of the scaling dimensions  via
the Bethe Ansatz  is given by special values of the parameters in the grand
canonical partition function

\begin{equation}
  \label{eq:1}
  G(K; \tau, p) = \sum_{n=0}^\infty K^n Z_n(\tau,p)
  .
\end{equation}
The location of this point is reported exactly in \cite{blote1989a-a}
\begin{align*}
  w &= K_{bn}^2 \tau_{bn} = \left\{ 2 - [1-2\sin(\theta/2)]
    [1+2\sin(\theta/2)]^2 \right\}^{-1} \\
  K_{bn} &= -4 w
  \sin(\theta/2) \cos(\pi/4 - \theta/4) \\
  p_{bn} K_{bn} &= w [1 + 2\sin(\theta/2)] \\
  \theta &= -\pi/4 \qquad \text{(branch 3 in \cite{blote1989a-a})}\; .
\end{align*}
Alternatively, this can be expressed in explicit algebraic numbers or evaluated numerically as $K_{bn} = 0.446933\ldots$, $p_{bn} = 0.275899\ldots$, and $\tau_{bn} = 2.630986\ldots$.

From the value of $K_{bn}$ we can give the conjectured value of the connective constant $\mu$ \eqref{eq:mu-def} in the canonical model (\ref{eq:canonical}) at fixed Boltzmann weights $(\tau,p)=(\tau_{BN},p_{BN})$ as
\begin{align}
  \mu(\tau_{bn},p_{bn}) = \frac{1}{K_{bn}} &= 
\left(
\frac{\sqrt{2}}2+
\frac{\sqrt{2}}2\sqrt{2-\sqrt{2}}+
\sqrt{2-\sqrt{2}}\right)
\sqrt{2-\sqrt{2-\sqrt{2}}}\nonumber\\
&=2.23746994\ldots
\end{align}
The set of scaling dimensions evaluated in \cite{warnaar1992b-a} at this (multi-)critical point are
\begin{equation}
  x_\ell = \frac{\ell^2}{16} - \frac{1}{6}\quad \mbox{ for } \ell\in\mathbb{N}.
\end{equation}
The thermal $x_t$ and magnetic $x_h$ scaling dimensions were identified \cite{warnaar1992b-a} as
\begin{equation}
  x_h = x_1 = -\frac{5}{48}= -0.10416\ldots \quad \mbox{ and } \quad x_t =x_2 = \frac{1}{12} = 0.08333\ldots\;.
\label{eqn:bn-scaling-dims}
\end{equation}
It can be seen that these scaling dimensions are not those of
unweighted self-avoiding walks; this is compatible with the
hypothesis that they are those of a collapse multi-critical point.

The exponents $\nu$ and $\gamma$ were then calculated \cite{warnaar1992b-a} in the standard
way as
\begin{align}
  \nu   &= \frac{1}{2- x_t}   = \frac{12}{23} = 0.52174 \ldots\\
  \gamma &= 2 \nu \,(1 - x_h) = \frac{53}{46} = 1.15217 \ldots\;.
\end{align}

\section{Simulation results}
\label{sec:results}

We simulated the model (\ref{eq:canonical}) at the fixed values of $p
= p_{bn}$ and $\tau = \tau_{bn}$ using the Pruned-Enriched Rosenbluth Method (PERM) \cite{Grassberger1997}. This method is based on the traditional Rosenbluth and Rosenbluth sampling method where biased samples of polymer configurations are generated along with a weight factor such that the weighted average over all polymer configurations will converge towards the correct Boltzmann average. PERM improves the efficiency of this algorithm by making multiple copies of partially grown chains that have a large statistical weight ('enriching') and discarding configurations with small statistical weight ('pruning'). 
We ran three simulations with maximal length $N_{max} = 1024$, $2048$ and $4096$, growing $S \simeq 10^{7}$ independent walks each and collecting from $8.7 \cdot 10^7$ to $1.5 \cdot 10^8$ samples at each maximal length. The number of samples adjusted by the number of their independent growth steps is between $2.1 \cdot 10^5$ and $4.8 \cdot 10^5$ ``effective samples''.

\begin{figure}[ht!]
  \centering
  \includegraphics{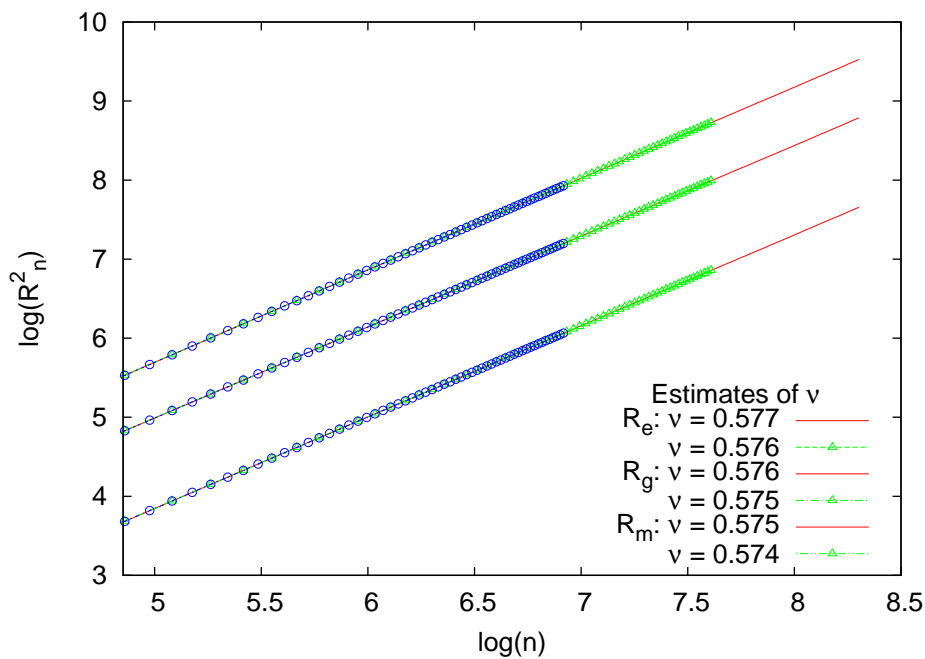}
  \caption{A double-logarithmic plot of three different measures
    $R_n^2$ of the polymer size, the mean-square end-to-end distance
    $\langle R_e^2\rangle_n$ (top), the square of the radius of
    gyration $\langle R_g^2\rangle_n$ (bottom), and the mean-square
    distance of a monomer to its end points $\langle R_m^2\rangle_n$
    (middle) versus the length $n$ of the path. The data from
    different simulations up to lengths $1024$ (blue circles), $2048$ (green triangles), 
    and $4096$ (red line) clearly overlap. The exponent estimates for $\nu$ shown are
    derived from fitting straight lines through lengths $1024$ to
    $2048$ and $2048$ to $4096$, respectively.}
\label{fig:R}
\end{figure}

In Figure~\ref{fig:R} we plot on a double-logarithmic scale the three
different measures $R_n^2$ of the polymer size. From various fits we
consistently find estimates of $\nu$ near $4/7$ rather than
$12/23$. Our best estimate is 

\begin{equation}
  \nu = 0.576(6)\;.
\label{eqn:nu-estimate}
\end{equation}
This leads to an estimate of the thermal scaling dimension as
\begin{equation}
  x_t = 0.26(2)\;.
  \label{eqn:xt-estimate}
\end{equation}
\begin{figure}[ht!]
  \centering
  \includegraphics{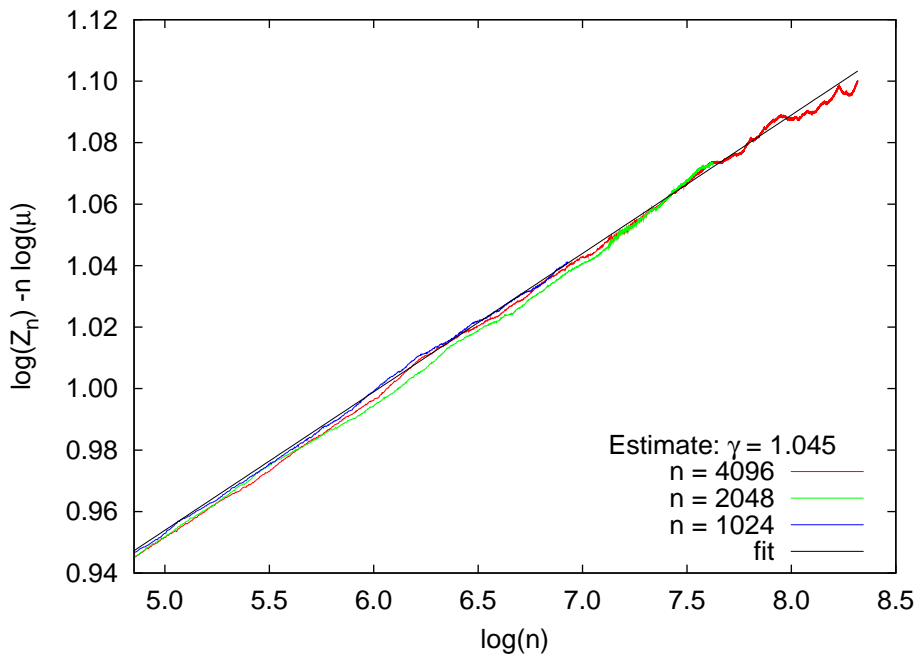}
  \caption{A double-logarithmic plot of the normalized partition
    function $Z_n/\mu^n$ versus the length $n$ of the path. The
    difference between data from different simulations up to lengths
    $1024$, $2048$, and $4096$ indicate the accuracy achieved. The
    exponent estimate for $\gamma$ shown is derived from fitting
    straight lines through lengths $256$ to $1024$ to data from the
    best-converged simulation.}
\label{fig:Z}
\end{figure}
To obtain an estimate of $\gamma$ we looked at
the scaling of the canonical partition function \eqref{eq:scaling-z}.
We first measured $\mu$ by a simple linear fit obtaining $2.2375(1)$, and observing that our value matches the
value obtained from the BN-model, we then assumed $\mu = \mu_{bn}$.  We
hence plotted, in Figure~\ref{fig:Z}, on a double-logarithmic scale the
normalized partition function $Z_n/\mu_{bn}^n$ versus the length $n$
of the path.

This allows us to estimate $\gamma$ from straight line fits which we
give as
\begin{equation}
  \gamma = 1.045(5)\; .
\label{eqn:gamma-estimate}
\end{equation}
Interestingly, this value is different from both the $\theta$-point
value of $8/7=1.14228\ldots$ and the BN-point $53/46=1.1521\ldots$.
Using $ \gamma = 2 \nu (1 - x_h)$ and our estimates of $\gamma$ and $\nu$ in equations~(\ref{eqn:gamma-estimate}) and (\ref{eqn:nu-estimate}) gives us the estimate
\begin{equation}
  x_h = 0.093(13)\;.
  \label{eqn:xh-estimate}
\end{equation}
We point out that this estimate is positive while the conjectured value above in equation~(\ref{eqn:bn-scaling-dims}) is not.

\begin{figure}[ht!]
  \centering
  \includegraphics{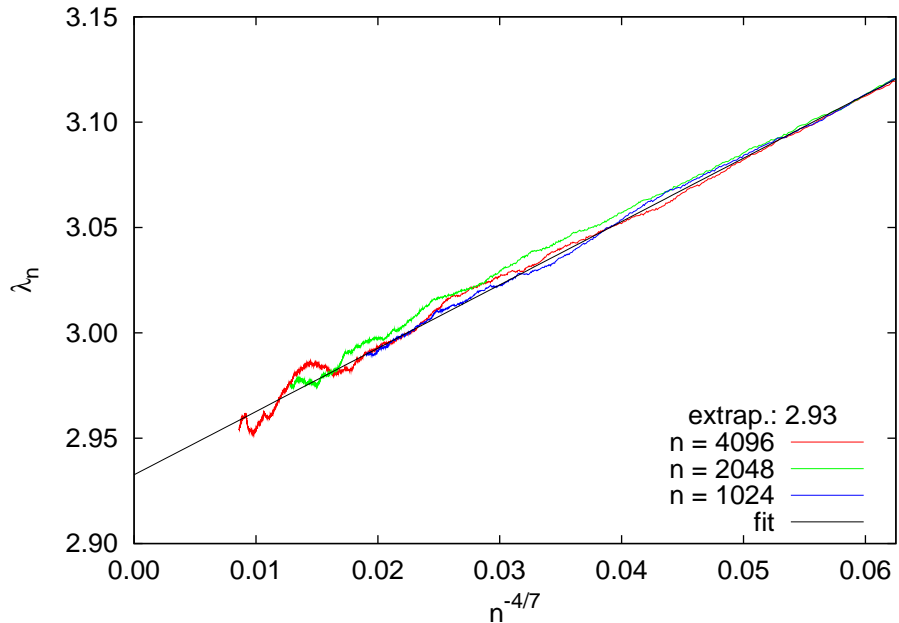}
  \caption{A plot of the finite-size amplitude ratio combination
    $\lambda_n$ versus $n^{-4/7}$, together with a straight-line
    fit. The difference between data from different simulations up to
    lengths $1024$, $2048$, and $4096$ indicate the accuracy
    achieved. The extrapolated value for $\lambda$ shown is derived
    from fitting straight lines through lengths $256$ to $1024$ to
    data from the best-converged simulation.}
\label{fig:lambda}
\end{figure}

To obtain an independent estimate of $x_h$ we attempted to estimate the universal quantity $\lambda$ described above. In Figure~\ref{fig:lambda} we plot the finite-size amplitude ratio
combination $\lambda_n=(4B_n-1)/(2A_n)$ versus $n^{-4/7}$,  which is the natural scale
given the results above for the size measures. We find an estimate of
this universal value as
\begin{equation}
  \lambda = 2.93(3)\; .
\end{equation}
Unfortunately, the error estimate here is relatively larger than that estimated from the partition function analysis and we estimated $x_h=0.12(4)$, which encompasses our more precise estimate in (\ref{eqn:xh-estimate}).

 For the sake of completeness we have also found estimates of the 
universal amplitude ratios $A$ and $B$ by fitting against a correction of $n^{-4/7}$, which provide consistent straight fit extrapolations
\begin{equation}
  A = 0.1534(10) \quad \text{and} \quad B = 0.475(5)\;.
\end{equation}
These values of $A$ and $B$ are different from the values for any of the three collapse models considered in \cite{owczarek1994b-:a}.

\section{Conclusions}
\label{sec:conclusions}

We have simulated the special point, known as the BN (Bl\"{o}te-Nienhuis) point,
of the semi-flexible VISAW model of polymer collapse, which is associated with
an integrable branch of the $O(n)$ loop model \cite{blote1989a-a}. The exponent
estimates we find, $\nu = 0.576(6)$ and $  \gamma = 1.045(5)$, are not in accord
with those previously found from the $O(n)$ loop model. Our estimate of $\nu$ is
compatible with the value accepted for the $\theta$-point, which is
$4/7=0.5714\ldots$, and in good agreement with the results of Foster and
Pinettes \cite{foster2012a-a} who have used transfer matrices and DMRG. However,
our estimate of $\gamma$ is not comparable to any known value. We have found
estimates of $\gamma$ via two different methods: one method we used involved the
direct estimation of the exponent from the partition function, and the other
used results from conformal field theory and universal amplitude ratios of
different size measures of the polymer; our estimates from these two methods
broadly agree.

Our results seem to suggest that the BN point is $\theta$ like, at least with respect to its size scaling exponent with $\nu = 4/7$.
This may seem at odds with our recent claim \cite{Bedini:2013kd} that the VISAWs, which don't weight straight segments and have $p = 1$ in the notation of this
paper, have a collapse transition in the same universality class as the
Interacting Self-Avoiding Trails \cite{Owczarek:2007dn}. This was based, however, upon analysis of the specific heat. Of course, the two claims are not in direct contradiction but they lie uncomfortably together. In particular, it
leaves open the question whether the conclusion that the ISAT universality
class extends down to, and importantly includes, the VISAW line, where the weight for crossings segments
($\tau_x$ in \cite{Bedini:2013kd}) vanishes. On the other hand, if both claims are true, there must be a change of universality class on varying $p$.  It should be emphasised however  that an estimate of the exponent $\nu$ for collapsing VISAWs is not
available at the moment, since the lack of knowledge of the exact location of
the transition for $p=1$ makes obtaining good estimates a significantly
harder task.

Clearly, something subtle is occurring in this system if our numerical
analysis is accurate. Of course, large corrections to scaling may  be at work
here. In any case, further theoretical work is needed to tease out this issue.

\ack

Financial support from the Australian Research Council via its support
for the Centre of Excellence for Mathematics and Statistics of Complex
Systems and through its Discovery program is gratefully acknowledged
by the authors. A L Owczarek thanks the School of Mathematical
Sciences, Queen Mary, University of London for hospitality.


\section*{References}

\begin{thebibliography}{10}

\bibitem{foster2012a-a}
D.~P. Foster and C.~Pinettes,
\newblock J. Phys. A. {\bf 45}, 505003 (2012).

\bibitem{nienhuis1982a-a}
B.~Nienhuis,
\newblock Phys. Rev. Lett. {\bf 49}, 1062 (1982).

\bibitem{duplantier1987a-a}
B.~Duplantier and H.~Saleur,
\newblock Phys. Rev. Lett. {\bf 59}, 539 (1987).

\bibitem{Vanderzande:1991bv}
C. Vanderzande, A.~L. Stella, and F. Seno,
\newblock Phys. Rev. Lett. {\bf 67}, 2757 (1991).

\bibitem{Stella:1993fz}
A.~L. Stella, F. Seno, and C. Vanderzande,
\newblock J Stat Phys {\bf 73}, 21 (1993).

\bibitem{Foster:1992gc}
D.~P. Foster, E. Orlandini, and M.~C. Tesi,
\newblock J. Phys. A. {\bf 25}, L1211 (1992).

\bibitem{blote1989a-a}
H.~W.~J. Bl{\"{o}}te and B.~Nienhuis,
\newblock J. Phys. A. {\bf 22}, 1415 (1989).

\bibitem{batchelor1989a-a}
M.~T. Batchelor, B.~Nienhuis, and S.~O. Warnaar.,
\newblock Phys. Rev. Lett. {\bf 62}, 2425 (1989).

\bibitem{nienhuis1990a-a}
B.~Nienhuis,
\newblock Int. J. Mod. Phys. B {\bf 4}, 929 (1990).

\bibitem{warnaar1992b-a}
S.~O. Warnaar, M.~T. Batchelor, and B.~Nienhuis,
\newblock J. Phys. A. {\bf 25}, 3077 (1992).

\bibitem{Bradley:1990eh}
R.~Bradley, Phys. Rev. A {\bf 41}, 914 (1990).

\bibitem{batchelor1993a-a}
M.~T. Batchelor,
\newblock J. Phys. A. {\bf 26}, 3733 (1993).

\bibitem{prellberg1994a-:a}
T.~Prellberg and A.~L. Owczarek,
\newblock J. Phys. A. {\bf 27}, 1811 (1994).

\bibitem{foster2003a-a}
D.~P. Foster and C.~Pinettes,
\newblock J. Phys. A {\bf 36}, 10279 (2003).

\bibitem{madras1988a-a}
N.~Madras and A.~D. Sokal,
\newblock J. Stat. Phys {\bf 50}, 109 (1988).

\bibitem{cardy1989a-a}
J.~L. Cardy and H.~Saleur,
\newblock J. Phys. A. {\bf 22}, L601 (1989).


\bibitem{caracciolo1990a-a}
S.~Caracciolo, A.~Pelissetto, and A.~D. Sokal,
\newblock J. Phys. A. {\bf 23}, L969 (1990).

\bibitem{owczarek1994b-:a}
A.~L. Owczarek, T.~Prellberg, D.~Bennett-Wood, and A.~J. Guttmann,
\newblock J. Phys. A {\bf 27}, L919 (1994).

\bibitem{Grassberger1997}
P. Grassberger,
\newblock Phys. Rev. E {\bf 56} 3682 (1997).

\bibitem{Bedini:2013kd}
A. Bedini, A.L. Owczarek, and T. Prellberg,
\newblock Phys. Rev. E {\bf 87}, 012142 (2013).

\bibitem{Owczarek:2007dn}
A.L. Owczarek and T. Prellberg,
\newblock Phys. A {\bf 373}, 433 (2007).

\end{thebibliography}
\end{document}